\documentclass{article}
\usepackage[utf8]{inputenc}
\usepackage{wordlike} 

\usepackage{hyperref}
\usepackage{listings}
\usepackage{amsmath}
\usepackage{graphicx}
\usepackage{float}

\usepackage{xspace}
\newcommand{\homos}{\textit{Homo sapiens }\xspace}

\newcommand{\mouse}{\textit{Mus musculus }\xspace}

\title{A word recurrence based algorithm to extract genomic dictionaries}
\author{Vincenzo Bonnici, Giuditta Franco, Vincenzo Manca}
\date{}

\begin{document}

\maketitle

\begin{abstract}
Genomes may be analyzed from an information viewpoint as very long strings, containing functional elements of variable length, which have been assembled by evolution. In this work an innovative information theory based algorithm is proposed, to extract significant (relatively small) dictionaries of genomic words. Namely, conceptual analyses are here combined with empirical studies, to open up a methodology for the extraction of variable length dictionaries from genomic sequences, based on the information content of some factors. Its application to human chromosomes highlights an original inter-chromosomal similarity in terms of factor distributions. 
%VB: considerando che ci rifacciamo alla rdd definita da altri in passato, magari direi che non sono proprio new theories. 
%GF Ho cambiato con approcci: la novita' dell'approccio sta nel fatto di cercare un vocabolario per il linguaggio genomico, con una combinazione di metodi in parte gia' noti
\end{abstract}

%In this perspective the relationship between a long string and its factors assumes particular importance, and genomic dictionaries (that is, collections of specific genomic portions) are defined and investigated as fingerprints (of given genomes) that exhibit properties of interest.

\section{Introduction}
\label{sec:intro}
Human genome computational analysis is one of the most important and intriguing research challenges we are currently facing. Genomes carry the main information underlying life of organisms and their evolution, including a system of molecular (reading, writing, and signal transmission) rules which orchestrate all cell functions. Most of these rules and especially the way they cooperate are unknown, while this being a problem of great scientific and medical interest~\cite{personal}, due mainly to currently incurable diffused genetic diseases. Our work here follows and outlines some trends of research which analyze and interpret genomic information, by assuming the genome to be a book encrypted in a language to decipher~(see for example~\cite{mant,searls,sad,nature4,booklet,Ling}). Namely, this analysis may be developed by sequence alignment-free methods based on information theoretical concepts, in order to convert the genomic information into a comprehensible mathematical form, such as a dictionary of variable-length factors that collects words of the unknown genomic language.

According to a common approach in computational genomics (e.g.,~\cite{zhang,ortuno2007keyword,encode2012integrated,zambelli2012motif,castellini}), a genome is represented by a string over the alphabet $\Gamma =\{A,C,T,G\}$, when secondary and ternary structure of DNA double helix are neglected, that is, a genome $G$ is an element of $\Gamma^\star$. This representation easily leads to affinities with a text, written in a natural language, which is comprehensible by means of its vocabulary, giving both syntax and semantic of {\it words}. Chomsky taught us that the elaboration of words (as sequences of symbols generated or recognized by a computational model) is crucial for formal languages, which paved the base of computer science, while Shannon gave birth to information theory working on codes (systems of words) equipped with a probability distribution. The concept (definition, characterization) of {\it word} is indeed central to understand the language in which information is organized within a text or a (genomic) string. 

%With an analogy with literary texts, genomic sequences are then composed by small unities, which represent the building blocks of a genomic language to discover (like it is with words of a natural language). We refer to them as genomic elements, genomic words, or genomic units, and our analysis is based on notions of dictionary (i.e., collection of words) and distribution (often defined on dictionaries), as well as on concepts derived from probability and information theory. 

An example of genomic dictionary may be found within the structure surrounding eukaryotic genes. The coding region of a gene is composed by a starting untranslated region ($5^\prime-UTR$), a specific starting codon (usually $ATG$), an interleaving of exons and introns followed by a termination codon ($TAG$, $TAA$ or $TGA$) and an untranslated tailing region ($3^\prime-UTR$). %Specific genomic words, such as $AGGAGGT$, have been recognized for ribosomal binding in both $5^\prime-UTR$ and $3^\prime-UTR$ regions, where in particular the polyadenylation signal $AATAAA$, poly-$A$s, and a general $AT$ content were identified. Moreover, $5^\prime$-flanking regions of genes contain specific Transcription Start Site (TSS) sequences, from where RNA polymerase starts the transcription, and genes are preceded by promoter regions within short distance, where TFBSs (transcription factor binding sites) $CAAT$-like and $TATA$-like boxes are well known components, and regulatory genomic elements (such as enhancers and silencers) displaced within long distances~[\cite{franco}].
However, in \homos genes cover a relatively small fraction of the entire genome, while the rest of it, first considered {\it junk DNA}, is either transcribed, into regulatory elements, or associated with some other biochemical activity: hence, it is covered by (generally long) functional elements. %Namely, it is not clear how dark matter communicates with both the close and the far genes it affects, and how millions of genetic switches control gene activation. 
%%%Our knowledge on the structure and function of the human genome has been strongly influenced by main results of the ENCODE (Encyclopedia of DNA elements) project~\cite{encode2012integrated}, which have changed our previous characterization of genetic regions. %Computational methodologies were applied to find genes' transcription factors as well as to identify the repetitive and recombinant nature of genomes~\cite{haubold2006repetitive,fraMil}. 
According to recent advancements, the concept of {\it functional element} is central, defined as a genomic segment that codes for a defined biochemical product or displays a reproducible biochemical signature~\cite{barcode,booklet}. Furthermore, the distinct distribution of transcribed RNA species across segments suggests that underlying biological activities are captured in some genome segmentation. 
%VB: attenzione perchè le parole che estraiamo non hanno queste proprietà di  biochemical product or displays a reproducible65biochemical signature
% Ho rigirato un po' i discorsi..
In such a context, catching most of the `significant' words (from these segments) which were naturally selected during evolution would be a first step to understand at least the syntax of an hypothetical genomic language, whose semantics may be possibly studied by the support of epigenomics. An information theory based analysis clearly plays an important role in deciphering such a language, and in the literature there are several examples regarding information theory applications to biological sequence analysis, for example reviewed in~\cite{vinga2013information}, that confirm the linkage between DNA fragments and their information content~\cite{zhang,sad,barcode,sims,chor,zheng}.

%A genome-wide (alignment-free) approach to investigate the genomic language is based on dictionaries to be discovered (i.e. suitably extracted by the genome string) by a reverse engineering process where biochemical and statistical protocols are applied. Successively, once peculiar features of genomic element families are extracted, computational approaches are applied to characterize them in the whole genome sequence or in sequences of other species. 

Also fixed length dictionaries show some interesting properties. Namely, in~\cite{carpena2009level,hackenberg2012clustering} the authors applied a methodology developed for literary text to extract fixed length genomic dictionaries. An analysis regarding the intersection of fixed length dictionaries coming from human chromosomes is reported in this paper. Examples of fixed length dictionary extraction procedures could be provided by applying notions such as word multiplicity or word length distributions. On the other hand, graphical investigative analyses, based on expected frequency gaps, show the unpredictable behaviour of genomic sequences and help to detect peculiar words~\cite{fraMil}. Following the terminology from our previous work~\cite{castellini}, given a genome $G$ we call $D_k(G) \subseteq \Gamma^k$ the {\it $k$-dictionary} of all $k$-mers occurring in the genome $G$.

If we think of a book, semantically significant words have a fairly medium number of occurrences (they are not over-represented, as conjunctions and prepositions, and only some of them are underrepresented, as signatures, neologisms, specialist words) and they are clustered according to the topic described in that part of the book. Analogously, it is clear why several works are focused on finding genomic words exhibiting some special kind of (somehow clustered) repetitiveness, with a global frequency quite different than the expected frequency in purely random sequences having the same length of an investigated genome~\cite{cover,holland,zhang,kong,chor,sims}. A very relevant and peculiar word periodicity is revealed by the {\it Recurrence Distance Distribution (RDD), which measures the frequency at which a given word occurs at given distances}~\cite{RTD} (for an application to real genomes, see~\cite{rdd}).

In this paper, we start from a modified version of the algorithm introduced in~\cite{carpena2009level}, in order to apply it to real genomes (e.g., human chromosomes). We call it V-algorithm, from the first name of the authors who designed it. Both these original and modified algorithms are aimed at finding words forming local clusters (the approach is explained in section~\ref{local}). Then we propose a new RDD-based algorithm, we call it W-algorithm, which extracts variable length dictionaries of interests from several real genomic sequences and collects words having a recurrence distribution maximally different than their random distribution. Such a selection is developed by computing the (locally) maximum divergence, from random sequences, of the RDD of each string obtained by elongating an initial {\it seed word} over the genome. The divergence from random sequences is a crucial issue in information analysis of strings~\cite{gatlin1972information,harter1974probabilistic} and in analyzing mathematical properties of dictionaries. The methodology in~\cite{carpena2009level} to find dictionaries is therefore here improved by the V-algorithm, and a more general approach is proposed (in section~\ref{global}) by means of the RDD based W-algorithm, that works with the global word recurrence distance distribution rather than with only a first slice of it.
 
Several studies from the state of the art define properties for words which result to be salient features in analysing genomic sequences~\cite{Makinen}. Minimal absent words, maximal or palindromic repeated words are some examples~\cite{garcia2011minimal,price2005novo,grissa2007crisprfinder}. Comparison of sequences for finding common substrings has been used for detecting protein domains via Markov's chains~\cite{bejerano2001markovian}. Other analyses recognize words that are statistically significant to compare two sequences~\cite{apostolico2010maximal}, or to discriminate sequence motifs~\cite{parida2014irredundant}. These approaches are focused on finding specific words to be used as key features of a string for analysing its property or for comparing it to another sequence~\cite{comin}. The extracted words are often sparsely located in the analysed sequence~\cite{sparse}, thus they do not constitute a real linguistic analysis of genomic strings. In~\cite{RTD}, on the other hand, an alignment-free distance measure (based on the return time distribution of k-mers) is employed for sequence clustering and phylogeny purposes. The approach presented in our study aims at extracting a set of words that represent, in a statistical way (that is, having a recurrence distribution maximally different than in a random sequence), the factors of the hidden language of a given genome. By analogy to linguistics, the extracted set of words constitute the dictionary of the unknown language. In fact, such dictionaries are shown to cover an high percentage of the sequence by (forcing) a minimal overlap of their occurrences.

In brief, we focus on two specific algorithms to extract genomic dictionaries with genomic words owning desired recurrence properties. Such automatically generated dictionaries were further $i)$ selected according to their genome coverage properties (that is, analysed in terms of contained words, their lengths, and their sequence and positional coverage over the source genome), $ii)$ biologically validated, $iii)$ filtered by elimination of infix and suffix words, and $iv)$ employed to cluster human chromosomes. Our methodology to extract and evaluate genomic dictionaries is explained in next three sections (where, respectively, the algorithmic approaches together with a dictionary validation criterion, the results obtained, and a discussion, are presented) and illustrated in Figure~\ref{meth}.

Specific software IGtools was developed in~\cite{tool} for extracting $k$-dictionaries, computing distributions and set-theoretic operations, evaluating empirical entropies and informational indexes. Both the V-algorithm and the W-algorithm were implemented within the {\it Infogenomics tools} framework, which is based on an engineered suffix array suitable for analyzing genomic sequences. The software is available also at \url{https://bitbucket.org/infogenomics/igtools/wiki/Home}. However, our work presented here is mainly a proof of concept, focused on the idea underlying the algorithms design, also supported by empirical results (namely on clustering human chromosomes), whereas algorithmic efficiency and implementation technology were not investigated. In this respect, advanced programming paradigm as the MapReduce could support our study with more computational analyses on genomic dictionaries~\cite{MapRed}, from both informational and linguistic viewpoints~\cite{Ling}. 
%For example, one may want to extract a set of words minimizing their lengths and maximizing their sequence coverage, but still preserving specific properties. 
%A common desired property is the possibility to reconstruct the source sequence by the extracted dictionary, especially in the field of genome sequencing and assembling where there are still several open problems ~[\cite{li2010novo,salzberg2012gage,schatz2012current,treangen2012repetitive,miyamoto2014performance}].

%Namely, in~[\cite{de1999combinatorics,colosimo2000special,carpi2001words,mignosi2002words,fici2006word,nando}] authors have studied the problem of factorizing a strings $S$ by a variable length dictionary $D$ (a dictionary that contains words having different lengths) such that, starting from $D$, $S$ can be unequivocally reconstructed, with a wished minimality in terms of factor lengths.

%The methodology is based on the notion of divergence between the real clustering coefficient of words and the empirical one in a random sequence. However, the original clustering coefficient approach~[\cite{carpena2009level}], used to discover key words in literary text, has been applied to real genomic sequences only later, in~[\cite{tesiVinc, hackenberg2012clustering}].

\section{Material and Methods} \label{tools}
 %We refer to words appearing once as {\it hapax}, to words appearing more than once (at least twice) as {\it repeats}, and to words which do not appear in the genome $G$ as {\it forbidden}~[\cite{hamp,algforb,mignosi2002words,fici2006word}]. Cardinalities of corresponding dictionaries $D_k(G)$, $H_k(G)$, $R_k(G)$, $F_k(G)$, respectively, have been computed and analyzed by varying both the word length $k$ and the genome G, as well as with respect to other more sophisticated informational indexes (such as $k$-lessicality, or $k$-dictionary selectivity, which take into account also the number of occurrences, i.e., {\it multiplicity}, of single repeats) in our previous works~[\cite{castellini,infobiotics,franco,fraMil}]. In particular, in~[\cite{nature,prof3}] the proper choice of the value $k$ was proposed for applying information theoretic concepts that express intrinsic aspects of genomes. The value $k = lg_2(n)$, where $n$ is the genome length, allows to define some indexes based on information entropies, helpful to find some informational laws (characterizing a general informational structure of genomes) and a new informational genome complexity measure. In~[\cite{nature}] this is computed by a generalized logistic map that balances entropic and anti-entropic components of genomes, that are related to their evolutionary dynamics. Specific software IGtools was developed in~[\cite{tool}] for extracting $k$-dictionaries, computing distributions and set-theoretic operations, evaluating empirical entropies and informational indexes.
 
RDD plays an important role in computational analysis of genomic sequences. Inspired by the fact the keywords are clustered in literary text, in~\cite{carpena2009level} RDD was used as basis in defining a clustering coefficient of words, while in~\cite{rdd} its application to coding regions shows the informational evidence of the codon language, and in~\cite{nair2005visualization,afreixo2009genome,bastos2011inter} some characterizations of recurrence behaviours were pointed out for very short $k$-mers. However, only fixed length dictionaries were extracted from real genomes by means of such a distribution~\cite{hackenberg2012clustering}.

In this section, we first summarize the genomic word extraction methodology reported in~\cite{carpena2009level}, which was our starting point to develop a variant of it, the V-algorithm, and then introduce a novel RDD-based extraction algorithm, called W-algorithm, by giving a special emphasis to issues regarding their applicability in extracting variable length dictionaries from human chromosomes. Here we propose also some criteria (based on suffix presence, biological relevance, and covering properties) to evaluate genomic dictionaries extracted by the W-algorithm, in order to optimize the whole methodology.

\subsection{A clustering coefficient C based approach} \label{local}
Authors of~\cite{carpena2009level} have used RDD to identify keywords by applying a methodology that associates a clustering coefficient $C$ to $k$-mers. The main idea is based on the fact that keywords are not uniformly distributed among a literary text, instead they are clustered. %The authors extended previous works introducing the idea that the word clustering must be statistically significant, namely not due to statistical fluctuation. 
Their approach combines the information provided by the spatial distribution of a word along the text (via the clustering coefficient) and its frequency, since the statistical fluctuation depends on the frequency. This basic approach has been used in~\cite{hackenberg2012clustering} to assign a relevance to $6$-mers and $8$-mers in \homos and \mouse. The $8$-mers were sorted by their normalized clustering coefficient (called $\sigma_{nor}$), and it has been shown that part of the top-200 clustered words (about $70\%$) appears in known functional biological elements, like coding regions and TFBSs.

The whole recurrence distribution is synthesised  with a single parameter $\sigma$, to quantify the clustering level, previously presented in~\cite{ortuno2007keyword} for studying the energy levels of quantum disorder systems~\cite{carpena2004new}, and a clustering degree $\sigma_{nor}$ assigned to words, for the identification of keywords in literary texts, obtained by means of the relation between the $\sigma$ of a real word and the theoretical expected one (coming from a theoretical hypothesized distribution), as in the following.

For a given word, the parameter $\sigma$ is the standard deviation of its normalized set of recurrence distances, $\sigma = s / \bar{d}$, where $s$ is the standard deviation of the recurrence distance distribution, and $\bar{d}$ is the average recurrence distance. When the RDD is a geometric distribution, the parameter is  denoted by $\sigma_{geo}$ and it is equal to $\sqrt{1-p}$, since $s = \sqrt{1-p}/p$ and $\bar{d} = 1/p$, where $p$ is the word frequency. Thus, the resultant normalized clustering measuring $\sigma_{nor}$ of the given word is given by $\frac{ \sigma}{ \sigma_{geo}} = \frac{ s / \bar{d}}{\sqrt{1-p}}$. For values of $\sigma_{nor}$ near to $1$, the recurrence distribution of the word is closed to the geometric one, thus it indicates a randomness of the word. In fact, a random sequence is generated by a Bernoullian process, then different occurrences of a given word are independent events, and the event of having $k$ occurrences of a word (in a segmentation unit) follows a Poisson distribution. Therefore, according to probability theory~\cite{feller2008introduction} its waiting time, that is the distance at which a word recurs, is an exponential distribution (having a geometric distribution as a discrete counterpart).

For words with low multiplicity, the statistical fluctuation are much larger, and its is possible to obtain an higher $\sigma_{nor}$ for rare words placed at random, and they would be misidentified as keywords.
Thus, the authors applied a correction by a Z-score measure that combines the clustering of a word and its multiplicity $n$. The resultant clustering measure $C$, is given by the following equation:
$C(\sigma_{nor}, n) = \frac{ \sigma_{nor} - \langle \sigma_{nor} \rangle(n) }{ sd(\sigma_{nor})(n) }$, where $\langle \sigma_{nor} \rangle(n) = \frac{2n -1}{2n+2}$ and $sd(\sigma_{nor})(n) = \frac{1}{\sqrt{n}(1+2.8n^{-0.865})}$. Parameter values were obtained via extensive simulations, by taking into account the distribution of $\sigma_{nor}$ in random texts. They represent the mean value and the standard deviation of such empirical distribution. The $C$ coefficient measures the deviation of $\sigma_{nor}$ with respect to the expected value in a random text, in units of the expected standard deviation. In this case, $C=0$ indicates randomness, $C>0$ that the word is clustered and $C<0$ that the word ``repels'' itself (it is generally distantiated from its previous occurrence).

In~\cite{carpena2009level} also an approach to explore the lineage of a word (from a short word to one of its possible elongations), without any knowledge about the effective word length, was provided. Given an initial word length $k_0$, some of the words in $D_{k_0}(G)$ are selected, according to their $C$ measure, that must be greater then a $C_0$ measure corresponding to a fixed percentile (usually $0.05$). Successively, for each of these initial word, their lineage is explored by selecting only the {\it elongations having a $C$ measure greater than $C_0$, and up to a fixed maximal word length}: these are properly the two points we changed in the V-algorithm presented in next section. In~\cite{carpena2009level}, the longest visited lineage is  selected as a word with semantic meaning, and the process is repeated for different values of $k_0$ (ranging in 2-35), until a dictionary is obtained by discarding repeating words.

%The V-algorithm still works on the elongation of short $k$-mers (called {\it seeds}) to longer words, but it seems more suitable for the application to real genomic sequences (see Table~\ref{tab:infa_extr_carp_carp1}). 

\subsection{The C based V-algorithm}

\begin{figure}[h]
\begin{center}
\includegraphics[width=14cm]{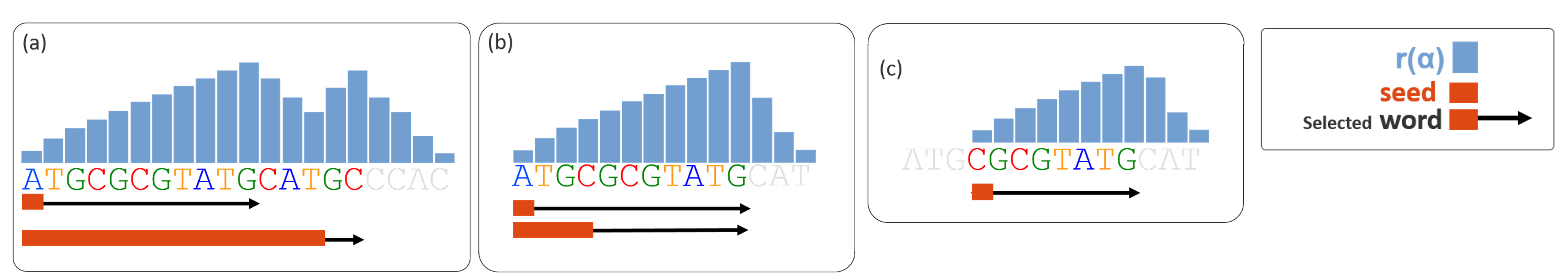}
\end{center}
\caption{\label{elong} {\bf Expansion procedure.} Word elongation is realized until the measure $r(\alpha)$ 
of the current word $\alpha$ does not decrease. In (b) the algorithm produces as an output the word ATGCGCGTATG, by starting from each seed, of length 1 and of length 4, while in (a) seeds of different lengths (1 and 14) allow to produce two different words (corresponding to the two peaks of $r(\alpha)$). 
In (c) a suffix of ATGCGCGTATG is generated, due to a different position of the seed long 1 with respect to (b).}
\end{figure}

The application of the above algorithm to wide genomic sequences comes with some issues. Indeed, in literary text the maximal length of words is known {\it a priori}, and parameters, such as the percentile threshold, can be calculated empirically. Working on genomes with the aim to discover an unknown genomic language requires a different approach. Although some words with biological meaning are already known, it would be a limitation to assign a value for the maximal word length in the dictionary we are extracting. Therefore, {\it the V-algorithm does not use any maximal word length threshold}. Moreover, word elongations are selected by comparing their $C$ measure with the one of their longest proper prefix, rather than with a prefixed $C_0$, and the elongated word (from length $k$ to $k+1$) is selected to be part of the output dictionary only if having a local maximum of the $C$-measure (see Figure~\ref{elong}).

% Carpena quindi tende a selezionare parole piu' lunghe, perche' si ferma prima di scendere sotto lo C0, quindi scollina, invece V-alg. rimane in cima alla montagna.

%VB: forse andrebbero specificati alpha e beta nella figura e non sono sicure che r(alpha), riportato in figura, sia già stato introdotto
%fatto tutto

In the above procedure we cannot exclude the fact that both a word  $\alpha\beta$ and its proper prefix $\alpha$ may own the desired properties. Indeed, there may exist significant words which are part of significant longer words, as illustrated in Figure~\ref{elong} (a), and to catch them it is fundamental to evaluate dictionaries achieved by starting from different seed lengths. By definition, if we run the elongation procedure by starting from the first symbol of $\alpha\beta$, then $\alpha$ is discovered as genomic word, and the elongation procedure stops without discovering the further elongation up to $\alpha\beta$. Hence, we run the elongation procedure over several sets of seed words, including also $\alpha x$ (to be evaluated as seed words, for $x \in \Gamma$) where $\alpha$ corresponds to the first peak of the parameter C, then to a genomic word selected to be an element of the dictionary we are extracting. Therefore, by our algorithms we allow roots (or prefixes) to be part of the genomic language, as it holds for natural languages.

As seed words we consider sets $D_{k_0}(G)$, with values of $k_0$ ranging from 1 to the {\it minimal forbidden length}, an informational index widely used in sequence analysis ~\cite{hamp,algforb,mignosi2002words,fici2006word}. %Another good choice could be the word length at which a relatively high percentage of {\it hapaxes} (also called {\it unique words}, as occurring once in the genome) compose the dictionary $D_k(G)$~\cite{castellini}. 
Our choice for seed lengths, in order to elongate 1-mers as well as longer seeds chosen according to a specific property of the analysed genome, is based on our previous works~\cite{castellini,fraMil,booklet}, where we have analyzed cardinalities of genomic dictionaries of $k$-mers, and informational indexes such as $k$-lessicality and $k$-dictionary selectivity, which take into account also the number of occurrences (i.e., {\it multiplicity}, of repeats) by varying the value of $k$. Namely, in~\cite{nature,prof3} the value $k = lg_2(n)$, where $n$ is the genome length, allowed us to define some indexes based on information entropies, helpful to find a new genome complexity measure. 

\subsection{The RDD based W-algorithm} \label{global}

Here we use RDD to calculate the divergence of the real distribution of a word within the genome from its frequency over a random string with the same genome length~\cite{kolmogorov,kong}. Such a divergence is used as a measure of the information content of a word. Low expressive words are elongated by an expansion procedure, until they reach a reasonable level of ``significance'' according to which they are classified as genomic words of the extracted dictionary.

\begin{figure}[htbp]
\vspace{-0.2cm}
\begin{center}
\includegraphics[width=12cm]{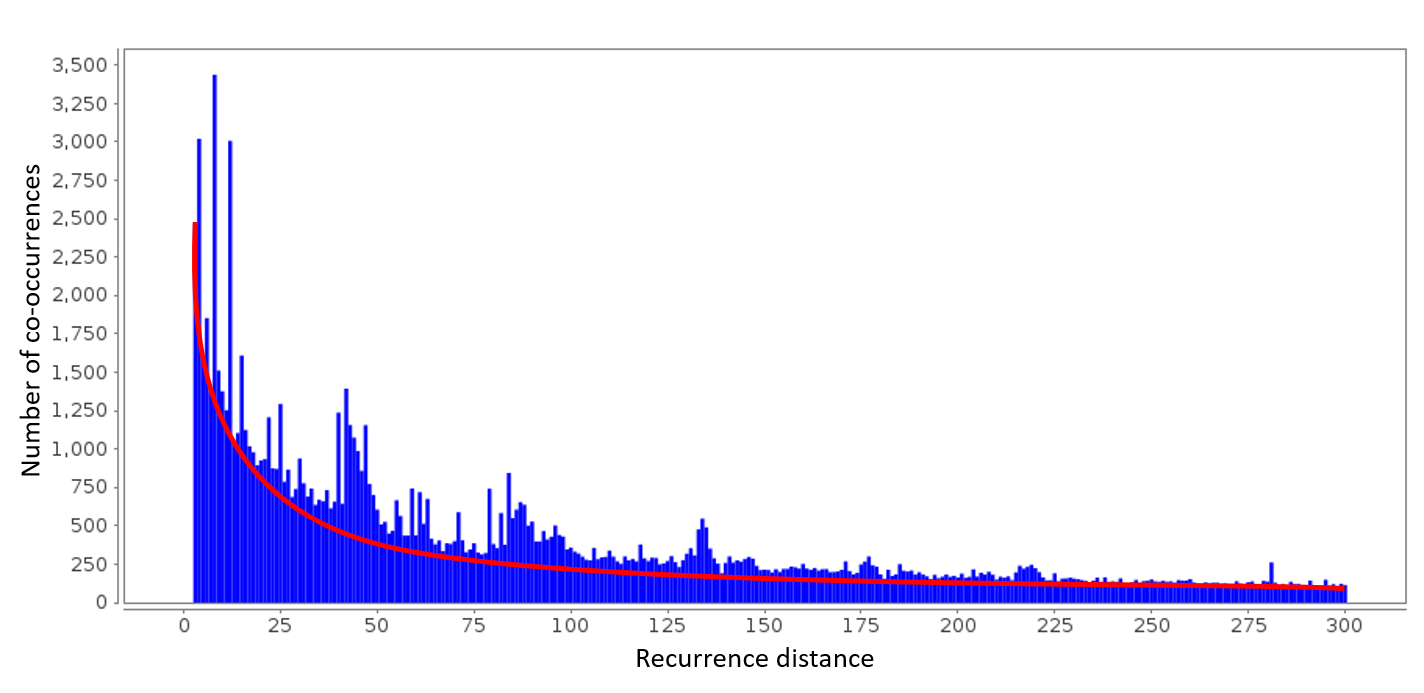}
%VB: il font in questa figura è troppo piccolo, ricordiamoci di rifarla o semplicemnte di scriverci sopra
\end{center}
\vspace{-0.5cm}
\caption{\label{rdd} RDD of word CGC (the jagged curve) in human chromosome 22, up to distance 300, with the ``best fitting'' exponential curve (the regular curve), which represents the waiting time of a Poisson distribution~\cite{feller2008introduction} ruling a word random occurrence.}
\end{figure}

We assume that the higher is the entropic divergence from the above exponential distribution, the more specialized and evolutionary selected is the genomic element. In this sense, low multiplicity words already represent elements owning high level of significance. %The underling idea indeed is that hapaxes are highly significant elements, because their probability to be selected among all the possible theoretical combinations is extremely low.
As for words with multiple occurrences (i.e., repeats), we associate their ``meaning'' with their repetitiveness-profile, as it is revealed by a ``good'' RDD. A good RDD means that a great number of recurrence distances have to occur, and the number of times such distances occur has to fall in a wide numeric interval. Roughly speaking, a repeat has to widely occur along the genomic sequence but it has also to show a reasonable level of specificity. In other terms, a word has to occur along the sequence several times and at different distances. See an example in Figure~\ref{rdd}, where the exponential distribution represents the random recurrence behaviour of the word. RDD of words along a real genomes is often sparse, meaning that several distances (of recurrence) actually do not appear in the genome. This is why we evaluate the sound (i.e., more fitting) exponential distribution after removing peaks, that are absent in exponential functions, and by imposing a normalization ensuring the overall unitary probability.

The degree of significance of a word to be selected for our dictionary is its {\it random deviation}, measured by the function in equation~(\ref{eq}), based on the the entropic divergence (such as the Kullback-Leibler divergence~\cite{cover}), between the real RDD of a word (over the analysed genome) and its expected exponential distribution.

More technically, given a word $\alpha$, which occurs in a genome $G$, we calculate its random deviation as the entropic divergence between its RDD and a suitable exponential distribution. To this aim, we first extract the real RDD of $\alpha$ over $G$, which we refer as $R_{\alpha}$.
Then we estimate a two parameters exponential distribution $E_{\alpha}$, by making use of the Nelder and Mead Simplex algorithm~\cite{nelder1965simplex}, a commonly used nonlinear optimization technique for problems for which derivatives may be not known.
A denoised distribution %VB: $R^prime_{\alpha}$  la definiamo e non la usiamo, quindi si può anche non definire come simbolo
is used as input for the estimation procedure: it is obtained by applying a low-pass filter (over $R_{\alpha}$) in order to attenuate peaks.
Afterwards we remove from $E_{\alpha}$ the domain values which are not present in $R_{\alpha}$, namely the gaps of $R_{\alpha}$.
Successively, both $R_{\alpha}$ and $E_{\alpha}$ are normalized in order to become probability distributions.
Finally, the random deviation of $\alpha$ is chosen as:
\begin{equation} 
\centering
r(\alpha) = \max(KL(R_{\alpha}, E_{\alpha})\text{ , } KL(E_{\alpha}, R_{\alpha})), \label{eq}
\end{equation}
where $KL$ is the asymmetric Kullback-Leibler entropic divergence. 

In our algorithm (reported in Listing 1) estimation of the information content of a word $\alpha$ is computed (at every step) by function $r(\alpha)$. Word elongation is realized until the (current) random deviation does not start to decrease. As it may be seen in Figure~\ref{elong}, smaller seeds allow the algorithm to generate words $\alpha$ corresponding to the first peak %VB: peak? Yes :)
(local maximum) of $r(\alpha)$. To produce a longer significant word $\alpha$, corresponding to the second peak of $r(\alpha)$, a longer seed has to be taken as a starting string. In all our computational experiments, $r(\alpha)$ showed to have only two peaks, whose localization depends on the genome length.

We would like to extract all the words $\alpha$ such that both $\alpha[1,|\alpha|-1]$ and $\alpha x $ (where $\alpha x$ is any elongation of $\alpha$ occurring in $G$ at least once) own a lower level of significance, namely a lower random deviation, with respect to $\alpha$. 
The goal can be reached by examining all the words within $G$ from monomers up to a word length equal to the maximum repeat length of $G$, and by discarding hapaxes. However, such an approach turns out highly expensive, and it cannot be applied efficiently for long genomes. Thus, we developed an expansion procedure with the aim of elongating seed words, let say monomers, up to more meaningful words. The (variable length dictionary) extraction algorithm, combining word elongation and random deviance test (in the expansion procedure) is given by two recursive functions in Listings 1 and 2, where $D_0$ denotes the set of seeds $D_{k_0} (G)$.

\begin{lstlisting}[caption={Extraction Algorithm},label=list:8-6,captionpos=t,float,abovecaptionskip=-\medskipamount, mathescape=true]
W:=$\emptyset$;
ForEach $\alpha \in D_0$:
    Elongate($\alpha,W$)
$W:= W \setminus D_0$;
Return $W$
\end{lstlisting}
%for i:=maxint to 0 do 
%begin 
%    j:=square(root(i));
%end;

\begin{lstlisting}[caption={Elongation procedure: Elongate ($\alpha, W$)},label=list:8-6,captionpos=t,float,abovecaptionskip=-\medskipamount, mathescape=true]
if $r(\alpha x) \leq r(\alpha), \forall x \in \Gamma$    then $W:= W \cup \{\alpha\}$
else ForEach  $x \in \Gamma$
         if $r(\alpha x) > r(\alpha)$ then Elongate($\alpha x$, $W$)
\end{lstlisting}

%\begin{algorithm}[h]
%\caption{ExtractFrom($G$,$D_0$)}\label{alg:extract_from}
%%$D_0 \gets D_k(G)$\\
%$W \gets \{\emptyset \}$\\
%\ForEach{ $\alpha \in D_0$ }{
%	Elongate($\alpha$, $W$)\\}
%$W \gets W \setminus D_0$\\
%\Return $W$
%\end{algorithm}

%\begin{algorithm}[h]
%\caption{Elongate($\alpha$,$W$)}\label{alg:elongate}
%\If{$r(\alpha x) \leq r(\alpha), \forall x \in \Gamma$}{
%$W \gets W \cup \{\alpha\}$\\
%}
%\Else{
%	\ForEach{$x \in \Gamma$}{
%		\If{$r(\alpha x) > r(\alpha)$}{
%			Elongate($\alpha x$, $W$)
%		}
%	}
%}

%\end{algorithm}

The main idea of the algorithm shown above is to compare the random deviation of a word with those of its elongations. If an elongation results in a word more significant than its root (i.e., its longest proper prefixes), then the root word is discarded and the elongated word is selected. The process is applied recursively over the word branching of the selected elements (see Listing 2). Seeds are discarded from the output dictionary. 

Three steps are implemented to compute random deviations. For all factors $\alpha$ of the genome $i)$ RDD of the current word $\alpha$ is computed, by also removing distribution noise (peaks) and transforming $R_\alpha$ into a probability distribution; $ii)$ an exponential distribution $E_\alpha$ is computed from $R_\alpha$ and normalized to be a probability distribution; $iii)$ random deviation $r_\alpha$ is computed by means of the Kullback-Leibler (entropic) divergence.

For the applications, we employ two elongating functions (along both different directions of the genome double string) and the resulting dictionary is the union of the dictionaries obtained with the two elongations. We refer with $W_{L2R}$ and $W_{R2L}$ as the dictionaries extracted by following the $5'-3'$ and $3'-5'$ verses, respectively, and with $W =  W_{L2R} \mathop{\cup} W_{R2L}$ as the resulting dictionary. Indeed, the concept of root expressed above is strictly related to the verse in which a text is written and read. However, DNA is a double helix where information resides on both strands, each one having an own reading verse. Therefore, the increasing random deviation must be investigated on either verses.

An application of the algorithm to real genomes shows interesting results, which are evaluated via information measurements on the extracted dictionaries, as described in next subsection.

\subsection{Dictionary evaluation}

Extracted dictionaries are evaluated by means of information measurements such as {\it the word length distribution} of their elements (see Tables~\ref{1WLD},~\ref{2WLD}), and its deviation from the minimal forbidden length of the genome. Two parameters we used to evaluate a dictionary D are the {\it sequence coverage}, which is the percentage of positions $i$ in the genome such that $G[j,k]$ is a word of the dictionary D for $j<i<k$, and the {\it average positional coverage}, which is the average over positions $i$ of number of words $G[j,k]$ for $j<i<k$ of the dictionary D. Intuitively, the first measures the portion of genome occupied by at least one word of the dictionary, and the second the average number of words from the dictionary that occupy each single position of the genome. They are denoted by $cov (G, D)$ and $avg (covp(G,D))$, respectively.

Dictionaries are tested in terms of both {\it sequence coverage} and {\it positional coverage} (see Tables~\ref{Seq},~\ref{Pos}). A ``good'' dictionary must have a high sequence coverage, but a low overlapping degree among its elements, too. In fact, if we consider $D_k(G)$ as a language, for a certain value of $k$, then it has the maximum sequence coverage (all positions of the genome would be involved by at least one $k$-mer) but also the maximum positional coverage, since each position of the sequence is involved by up to $k$ different words of the dictionary.

%number of words of D (counted with their multiplicity in the genome) that occupy each single position of the genome (average over positions $i$ such that $G[j,k]$ is a word of the dictionary D, for $j<i<k$). %that "are occupied" by at least one word of the dictionary D (i.e., the number of positions $i$, such that $G[j,k]$ is a word of the dictionary D, for $j<i<k$) and the {\it average positional coverage}, which is the average number of words of D (counted with their multiplicity in the genome) that occupy each single position of the genome (average over positions $i$ such that $G[j,k]$ is a word of the dictionary D, for $j<i<k$). They are denoted by $cov (G, D)$ and $avg (covp(G,D))$, respectively.

On an ideally good dictionary, both parameters are close to one, meaning that its words cover almost the entire genome and tend to not overlap. These parameters were computed both on single and union dictionaries, and on sub-dictionaries $W_k$ of fixed length words, in order to focus our analysis on groups of words having $avg (cov_p(G,W_k))$ as close as possible to one.

We also checked that our dictionaries are informative enough to contain most of the biologically annotated sequences, and have filtered them by elimination of suffixes and infixes in order to provide more significant languages. Moreover, by intersection of suitable extracted dictionaries from all human chromosomes we have developed a clustering which confirmed known similarities among them.

\section{Results} \label{main}
Both the algorithms described in previous section were run over all human chromosomes, by taking them into account individually, and then merging the $24$ sets into a single one. Analyses were run over the human reference assembly hg19 ( \url{https://www.ncbi.nlm.nih.gov/assembly/GCF_000001405.13/}). %The initial values (for the seed length) $k_0$ were taken in the range $5-12$. 

\subsection{Dictionaries extracted by the V-algorithm} \label{Valg}
Table~\ref{tab:infa_extr_carp_carp1} shows the number of extracted words (that is, dictionary sizes), for each single human chromosome, and their union at the bottom, for both the algorithm in~\cite{carpena2009level} and the V-algorithm, by starting from different seed lengths, and by implementing two filters as redundancy strategies: one discarding duplicates (same words coming from different seed lengths) and the other discarding prefixes (in order to estimate the relative amount of prefixes).
\begin{table}[ht]
\vspace{-0.4cm} \begin{center} 
\scalebox{0.7}{\begin{tabular}{|c|c|r|r|r|c|r|r|r|} \scriptsize
%\hline
Chr	 & 	& Original	 & 	Original  &  ratio & & V-algorithm  & 	V-algorithm & ratio \\
 & & no duplicates & no prefixes & & & no duplicates & no prefixes & \\\hline
1	 & 	 & 	276,178	 & 	210,728	 & 	0.763		 & 	 & 	57,064	 & 	57,055	 & 	1.000	 \\
2	 & 	 & 	281,698	 & 	227,544	 & 	0.808		 & 	 & 	119,582	 & 	118,368	 & 	0.990	 \\
3	 & 	 & 	259,805	 & 	203,888	 & 	0.785		 & 	 & 	102,640	 & 	101,142	 & 	0.985	 \\
4	 & 	 & 	251,067	 & 	201,760	 & 	0.804		 & 	 & 	108,229	 & 	106,879	 & 	0.988	 \\
5	 & 	 & 	259,167	 & 	207,300	 & 	0.800		 & 	 & 	112,846	 & 	111,581	 & 	0.989	 \\
6	 & 	 & 	255,025	 & 	198,487	 & 	0.778		 & 	 & 	106,193	 & 	104,510	 & 	0.984	 \\
7	 & 	 & 	269,392	 & 	208,465	 & 	0.774		 & 	 & 	113,139	 & 	111,840	 & 	0.989	 \\
8	 & 	 & 	259,586	 & 	206,241	 & 	0.794		 & 	 & 	118,551	 & 	117,295	 & 	0.989	 \\
9	 & 	 & 	212,362	 & 	152,523	 & 	0.718		 & 	 & 	33,886	 & 	33,878	 & 	1.000	 \\
10	 & 	 & 	234,663	 & 	186,844	 & 	0.796		 & 	 & 	100,616	 & 	99,595	 & 	0.990	 \\
11	 & 	 & 	249,374	 & 	188,012	 & 	0.754		 & 	 & 	94,484	 & 	93,417	 & 	0.989	 \\
12	 & 	 & 	247,842	 & 	187,931	 & 	0.758		 & 	 & 	99,147	 & 	97,579	 & 	0.984	 \\
13	 & 	 & 	176,546	 & 	149,563	 & 	0.847		 & 	 & 	81,634	 & 	78,868	 & 	0.966	 \\
14	 & 	 & 	209,881	 & 	162,515	 & 	0.774		 & 	 & 	94,312	 & 	90,313	 & 	0.958	 \\
15	 & 	 & 	207,173	 & 	177,125	 & 	0.855		 & 	 & 	107,114	 & 	103,917	 & 	0.970	 \\
16	 & 	 & 	229,208	 & 	166,653	 & 	0.727		 & 	 & 	62,732	 & 	62,673	 & 	0.999	 \\
17	 & 	 & 	204,905	 & 	160,475	 & 	0.783		 & 	 & 	85,091	 & 	84,303	 & 	0.991	 \\
18	 & 	 & 	161,710	 & 	131,900	 & 	0.816		 & 	 & 	65,985	 & 	65,558	 & 	0.994	 \\
19	 & 	 & 	258,781	 & 	197,822	 & 	0.764		 & 	 & 	123,913	 & 	122,541	 & 	0.989	 \\
20	 & 	 & 	171,474	 & 	131,434	 & 	0.766		 & 	 & 	66,320	 & 	65,597	 & 	0.989	 \\
21	 & 	 & 	130,763	 & 	100,427	 & 	0.768		 & 	 & 	50,698	 & 	50,233	 & 	0.991	 \\
22	 & 	 & 	147,002	 & 	120,259	 & 	0.818		 & 	 & 	77,797	 & 	74,511	 & 	0.958	 \\
X	 & 	 & 	279,938	 & 	213,093	 & 	0.761		 & 	 & 	124,793	 & 	123,006	 & 	0.986	 \\
Y	 & 	 & 	194,014	 & 	137,284	 & 	0.708		 & 	 & 	66,088	 & 	65,986	 & 	0.998	 \\\hline
union	 & 	 & 	4,281,701	 & 	3,737,766	 & 	0.873		 & 	 & 	1,813,776	 & 	1,798,241	 & 	0.991	 \\
\hline
\end{tabular}}
\end{center} \vspace{-0.2cm}
\caption{\label{tab:infa_extr_carp_carp1}
Number of extracted words for each human chromosome and the entire set union, both for original and modified clustering based algorithms. Ratios indicate the percentage of single sets, obtained by removing prefixes and duplicates, coming from seed length values in the range $5-12$.}
\end{table}

The result is that the V-algorithm is able to select a smaller set of words, with a lower gap between the two redundancy discarding strategies. 
This is essentially due to the fact that the higher is $k$ the lower are the $C$ measures of $k$-mers. Therefore, comparing the $C$ measure of a word, relatively longer than $k_0$, with the measure of its proper prefix is more restrictive than a comparison with the measure of the initial word of length $k_0$. From this behaviour, we can speculate that the V-algorithm selects words with an higher semantic meaning. 

In Table~\ref{tab:infa_extr_carp_carp1} it is evident that the V-algorithm extracts a smaller amount of duplicates and prefixes than the algorithm in~\cite{carpena2009level} (even when starting from seeds with different length). Indeed, {\it smaller variable length dictionaries were extracted} by the V-algorithm, with fewer duplicate discarding steps, and a smaller amount of prefixes (which needed to be discarded in the original algorithm).

\subsection{Dictionaries extracted by the W-algorithm} \label{Walg}
The RDD-based W-algorithm was applied (with values for seed length from the range $1-12$) to extract genomic dictionaries from each human chromosome, and some analysis were performed also on the union of such 24 dictionaries. However, here we show data only for some (more explicable) chromosomes, for (more significant) seed lengths up to~8.
\begin{table}[htbp] 
\vspace{-0.2cm} 
\begin{center}\scalebox{0.7}{
\begin{tabular}{r|r|r|r|r|r|r|r|r} \scriptsize
 & \multicolumn{8}{|c}{$k_0$} \\\hline
$k$	&	1	&	2	&	3	&	4	&	5	&	6	&	7	&	8	\\\hline
4	&	2	&	13	&	20	&		&		&		&		&		\\
5	&	31	&	134	&	202	&	272	&		&		&		&		\\
6	&	{\bf 63}	&	{\bf 349}	&	{\bf 517}	&	{\bf 995}	&	{\bf 1,261}	&		&		&		\\
7	&	57	&	180	&	232	&	350	&	475	&	1,343	&		&		\\
8	&	57	&	193	&	277	&	430	&	679	&	3,001	&	10,668	&		\\
9	&	10	&	144	&	241	&	529	&	1,073	&	7,602	&	29,521	&	53,314	\\
10	&	5	&	{\bf 201}	&	{\bf 326}	&	{\bf 794}	&	{\bf 1,391}	&	{\bf 9,126}	& 59,951	&	129,872	\\
11	&	2	&	151	&	233	&	569	&	923	&	4,302	&	{\bf 63,089}	&	{\bf 184,296}	\\
12	&		&	64	&	91	&	198	&	323	&	973	&	24,275	&	97,646	\\
13	&		&	21	&	30	&	51	&	81	&	225	&	4,592	&	20,670	\\
14	&		&	2	&	3	&	10	&	18	&	40	&	875	&	3,525	\\
15	&		&	2	&	2	&	5	&	6	&	11	&	190	&	724	\\
16	&		&	4	&	5	&	5	&	5	&	9	&	54	&	165	\\
17	&		&	1	&	1	&	2	&	2	&	3	&	17	&	54	\\
18	&		&		&		&		&		&		&	5	&	19	\\
19	&		&		&		&		&		&		&		&	5	\\
20	&		&		&		&		&		&		&		&	6	\\
21	&		&		&		&		&		&		&		&	3	\\
22	&		&		&		&		&		&		&		&	6	\\
23	&		&		&		&		&		&		&		&	1	\\
\hline
\end{tabular}} \end{center} \vspace{-0.2cm}
\caption{\label{1WLD} {\bf Word Length Distribution of human chromosome 1.} RDD-algorithm run on chromosome 1 produces dictionaries here partitioned according to the word length. Seed length is given as $k_0$ value, extracted word length as $k$. Cardinality of fixed length extracted words shows a bimodal trend for $k$-mers shorter than 7 (bold numbers are local maximum values). The longest word extracted by the algorithm is 24.}\vspace{-0.5cm}
\end{table}

\begin{table} [htbp] 
\begin{center} \scalebox{0.7}{
\begin{tabular}{r|r|r|r|r|r|r|r|r} %\scriptsize
 & \multicolumn{8}{|c}{$k_0$} \\ 
 \hline
$k$	&	1	&	2	&	3	&	4	&	5	&	6	&	7	&	8	\\
4	&		&	5	&	5	&		&		&		&		&		\\
5	&	17	&	54	&	108	&	179	&		&		&		&		\\
6	&	41	&	305	&	666	&	1,306	&	1,666	&		&		&		\\
7	&	{\bf 92}	&	{\bf 337}	&	{\bf 616}	&	{\bf 1,478}	&	{\bf 2,310}	&	{\bf 2,925}	&		&		\\
8	&	79	&	178	&	280	&	468	&	593	&	1,474	&	4,151	&		\\
9	&	43	&	142	&	248	&	562	&	811	&	3,879	&	14,614	&	39,347	\\
10	&	8	&	{\bf 221}	&	{\bf 542}	&	{\bf 1,325}	&	{\bf 2,140}	&	{\bf 9,106}	& 48,112	& 144,355	\\
11	&	{\bf 13}	&	197	&	479	&	1,284	&	2,115	&	6,986	&	{\bf 50,442}	&	224,644	\\
12	&		&	122	&	297	&	838	&	1,363	&	2,201	&	24,687	&	{\bf 303,163}	\\
13	&	2	&	53	&	119	&	327	&	579	&	774	&	6,403	&	136,135	\\
14	&	2	&	19	&	36	&	80	&	145	&	194	&	1,094	&	20,805	\\
15	&	2	&	7	&	9	&	21	&	33	&	50	&	291	&	4,193	\\
16	&		&	5	&	7	&	12	&	17	&	24	&	99	&	1,196	\\
17	&		&	2	&	3	&	5	&	6	&	9	&	27	&	327	\\
18	&		&	1	&	1	&	1	&	1	&	4	&	12	&	128	\\
19	&		&		&		&		&		&		&	2	&	43	\\
20	&		&		&		&		&		&		&	2	&	15	\\
21	&		&		&		&		&		&		&		&	6	\\
22	&		&		&		&		&		&		&		&	1	\\
23	&		&		&		&		&		&		&		&		\\
24	&		&		&		&		&		&		&		&	1	\\
\hline
 \end{tabular}} \end{center} \caption{\label{2WLD} {\bf Word Length Distribution of human chromosome 22.} RDD-algorithm run on chromosome 22 produces dictionaries here partitioned according to the word length. Seed length is given as $k_0$, extracted word length as $k$. A bimodal trend for $k$-mers shorter than 7 may be visualized, with local maximum values in bold. The longest word extracted by the algorithm is 23.}  \vspace{-0.8cm}
 % Slide 15.
\end{table}

The word length distribution (WLD) related to human chromosomes 1 and 22 are shown respectively in Table~\ref{1WLD} and Table~\ref{2WLD}, by reporting the cardinality of words both having a given length and being generated by the W-algorithm starting from a given seed length. A common feature (also for the other chromosomes) is to have {\it two modes} in the $k$-dictionary sizes, that is, two local maximum values (indicated in bold) for some lengths $k$. In Table~\ref{1WLD} such values are 6 (for seeds long from 1 to 5) and 10-11 (for seeds long from 2 to 8) , while in Table~\ref{2WLD} these values are 7 (for seed lengths from 1 to 6) and 10-11 (for seed lengths from 2 to 7). Although they have not fixed values, they are not very variable, if we consider that chromosome 22 and 1 are respectively the shortest and the longest ones (see Table~\ref{tab:infa_extr_carp_carp1}, where chromosomes are ordered by length). Let us here remark that 9 is the minimal forbidden length for all chromosomes, and 6 is a word length $k$ such that human chromosomes own all the possible $k$-mers (i.e., examers), with high multiplicity. In {\it E. coli}, as an example, the two modes have values 5 and 9.

Another empirical result, which was confirmed on all the other chromosomes, is that {\it the dictionary generated by starting from seeds long $k-1$ is a proper subset of that generated by starting from seeds long $k$, apart of the words long $k$}. In fact, words with the same length of the seed are eliminated by the algorithm and do not appear in the WLD tables. In other terms, numbers in one column of Table~\ref{1WLD} count some of the words counted on the same row in the next column (only along columns index greater than the seed length, due to the seed discarding policy of the algorithm). 

As described in section of Material and Methods, the extracted dictionaries are evaluated according to both their sequence and their (average) positional coverage: these data related to chromosome~1 are reported in Table~\ref{Seq} and Table~\ref{Pos} respectively, where it is clear that parameter goodness does not increase with the word or seed length $k_0$. 
\begin{table}[htbp]
 \vspace{-0.5cm} \begin{center} \scalebox{0.7}{
\begin{tabular}{r|r|r|r|r|r|r|r|r}%\scriptsize
 & \multicolumn{8}{|c}{$k_0$} \\\hline
$k$	&	1	&	2	&	3	&	4	&	5	&	6	&	7	&	8	\\\hline
4	&		&	0.0291	&	0.0291	&		&		&		&		&		\\
5	&	0.0309	&	0.0790	&	0.1362	&	0.1681	&		&		&		&		\\
6	&	0.0269	&	{\bf 0.3149}	&	{\bf 0.5504}	&	{\bf 0.7767}	&	{\bf 0.8426}	&		&		&		\\
7	&	{\bf 0.0742}	&	0.2479	&	0.3878	&	0.6430	&	0.7691	&	{\bf 0.8141}	&		&		\\
8	&	0.0285	&	0.0616	&	0.0899	&	0.1187	&	0.1384	&	0.1643	&	{\bf 0.2634}	&		\\
9	&	0.0115	&	0.0209	&	0.0303	&	0.0499	&	0.0615	&	0.0714	&	0.1593	&	{\bf 0.6315}	\\
10	&	0.0008	&	0.0054	&	0.0071	&	0.0128	&	0.0206	&	0.0329	&	0.0974	&	0.5388	\\
11	&	{\bf 0.0025}	&	{\bf 0.0077}	&	{\bf 0.0088}	&	{\bf 0.0108}	&	{\bf 0.0127}	&	0.0174	&	0.0602	&	0.3509	\\
12	&		&	0.0028	&	0.0031	&	0.0081	&	0.0089	&	0.0101	&	0.0342	&	0.2858	\\
13	&	0.0000	&	0.0006	&	0.0013	&	0.0054	&	0.0065	&	0.0070	&	0.0155	&	0.1209	\\
14	&	0.0035	&	0.0048	&	0.0049	&	0.0056	&	0.0065	&	0.0066	&	0.0101	&	0.0451	\\
15	&	0.0026	&	0.0036	&	0.0036	&	0.0050	&	0.0052	&	0.0052	&	0.0065	&	0.2140	\\
16	&		&	0.0016	&	0.0017	&	0.0017	&	0.0071	&	0.0028	&	0.0032	&	0.0090	\\
17	&		&	0.0011	&	0.0011	&	0.0012	&	0.0013	&	0.0013	&	0.0014	&	0.0031	\\
18	&		&	0.0006	&	0.0006	&	0.0006	&	0.0006	&	0.0012	&	0.0012	&	0.0020	\\
19	&		&		&		&		&		&		&	0.0000	&	0.0003	\\
20	&		&		&		&		&		&		&	0.0000	&	0.0002	\\
21	&		&		&		&		&		&		&		&	0.0001	\\
22	&		&		&		&		&		&		&		&	0.0000	\\
23	&		&		&		&		&		&		&		&		\\
24	&		&		&		&		&		&		&		&	0.0000	\\
\hline
\end{tabular}} \end{center} \vspace{-0.2cm}
\caption{\label{Seq} {\bf Human chromosome 1: sequence coverage values.} RDD-algorithm run on human chromosome 1 produces dictionaries, here partitioned according to the word length $k$ and the seed length $k_0$. Values in the table describe the portion of the genomic sequence covered by the dictionary, while bold numbers being the local maxima in the column (showing a bimodal trend).  \vspace{-0.9cm}}
\end{table}

By observing data in Table~\ref{Seq}, best coverage of the chromosome (corresponding value 0.84) is obtained by the examers obtained starting from 5-mers as seeds, while the average positional coverage of such a dictionary is 2.7715 (see Table~\ref{Pos}), which is far from one. However, this dictionary was our choice for the chromosome clustering analysis described below, because we gave a priority of importance to sequence coverage. Relatively to only positional coverage values, in Table~\ref{Pos} we may notice that words of length 10 (or longer, for instance 15) exhibit good (i.e., less than 2) values for any seed length up to 7, while examers have good positional coverage with shorter seeds (long up to 3).

By observing data in Table~\ref{Seq}, best coverage of the chromosome (corresponding value 0.84) is obtained by the examers obtained starting from 5-mers as seeds, while the average positional coverage of such a dictionary is 2.7715 (see Table~\ref{Pos}), which is far from one. However, this dictionary was our choice for the chromosome clustering analysis described below, because we gave a priority of importance to sequence coverage. Relatively to only positional coverage values, in Table~\ref{Pos} we may notice that words of length 10 (or longer, for instance 15) exhibit good (i.e., less than 2) values for any seed length up to 7, while examers have good positional coverage with shorter seeds (long up to 3).
\begin{table} [htbp]
 \vspace{-0.5cm} 
 \begin{center} \scalebox{0.7}{
\begin{tabular}{r|r|r|r|r|r|r|r|r}
 & \multicolumn{8}{|c}{$k_0$} \\\hline
$k$	&	1	&	2	&	3	&	4	&	5	&	6	&	7	&	8	\\\hline
4	&		&	1.0078	&	1.0078	&		&		&		&		&		\\
5	&	1.0807	&	1.1690	&	1.2411	&	1.4198	&		&		&		&		\\
6	&	{\bf 1.1539}	&	{\bf 1.3022}	&	{\bf 1.6590}	&	{\bf 2.3201}	&	{\bf 2.7715}	&		&		&		\\
7	&	1.0934	&	1.2876	&	1.4587	&	1.9817	&	2.5877	&	2.9160	&		&		\\
8	&	1.1569	&	1.2590	&	1.3125	&	1.4228	&	1.5184	&	1.5836	&	1.5572	&		\\
9	&	{\bf 1.4480}	&	{\bf 1.5411}	&	{\bf 1.5211}	&	{\bf 1.7039}	&	{\bf 1.8791}	&	{\bf 1.8661}	&	1.5470	&	1.7484	\\
10	&	1.0006	&	1.1090	&	1.1033	&	1.1697	&	1.1926	&	1.2632	&	1.2580	&	1.5457	\\
11	&	{\bf 4.0810}	&	{\bf 2.1729}	&	{\bf 2.0809}	&	{\bf 1.9100}	&	{\bf 1.7829}	&	{\bf 1.6131}	&	{\bf 1.3009}	&	1.3658	\\
12	&		&	1.0654	&	1.0624	&	1.1926	&	1.1809	&	1.1716	&	1.1507	&	1.3455	\\
13	&	1.0000	&	1.0000	&	1.0000	&	1.1355	&	1.3769	&	1.3530	&	1.2340	&	1.3709	\\
14	&	1.0000	&	1.0000	&	1.0000	&	1.0551	&	1.2244	&	1.2235	&	1.1687	&	1.3807	\\
15	&	1.000	&	1.1446	&	1.1445	&	1.1065	&	1.1739	&	1.1725	&	1.1444	&	1.2559	\\
16	&		&	1.2684	&	1.2636	&	1.2588	&	1.2539	&	1.1544	&	1.1447	&	1.1148	\\
17	&		&	1.0000	&	1.0000	&	{\bf 1.3982}	&	{\bf 1.3957}	&	{\bf 1.3948}	&	{\bf 1.3608}	&	{\bf 1.3440}	\\
18	&		&	1.0000	&	1.0000	&	1.0000	&	1.0000	&	1.0000	&	1.0015	&	1.0187	\\
19	&		&		&		&		&		&		&	1.0000	&	1.0000	\\
20	&		&		&		&		&		&		&	1.0000	&	1.0000	\\
21	&		&		&		&		&		&		&		&	1.0000	\\
22	&		&		&		&		&		&		&		&	1.0000	\\
23	&		&		&		&		&		&		&		&		\\
24	&		&		&		&		&		&		&		&	1.0000	\\
\hline
\end{tabular}}\end{center}  \vspace{-0.2cm}
\caption{ \label{Pos} {\bf Human chromosome 1: average positional coverage.} RDD-algorithm run on human chromosome 1 produces dictionaries, here partitioned according to the word length $k$ and the seed length $k_0$. Values in the table describe the average number of words (counted in the dictionary with their multiplicity on the chromosome) covering single positions, while bold numbers being the local maxima in the column.} \vspace{-0.9cm}
\end{table}

\begin{figure}[htbp] \vspace{-0.7cm} \centering
\includegraphics[width=0.5\linewidth]{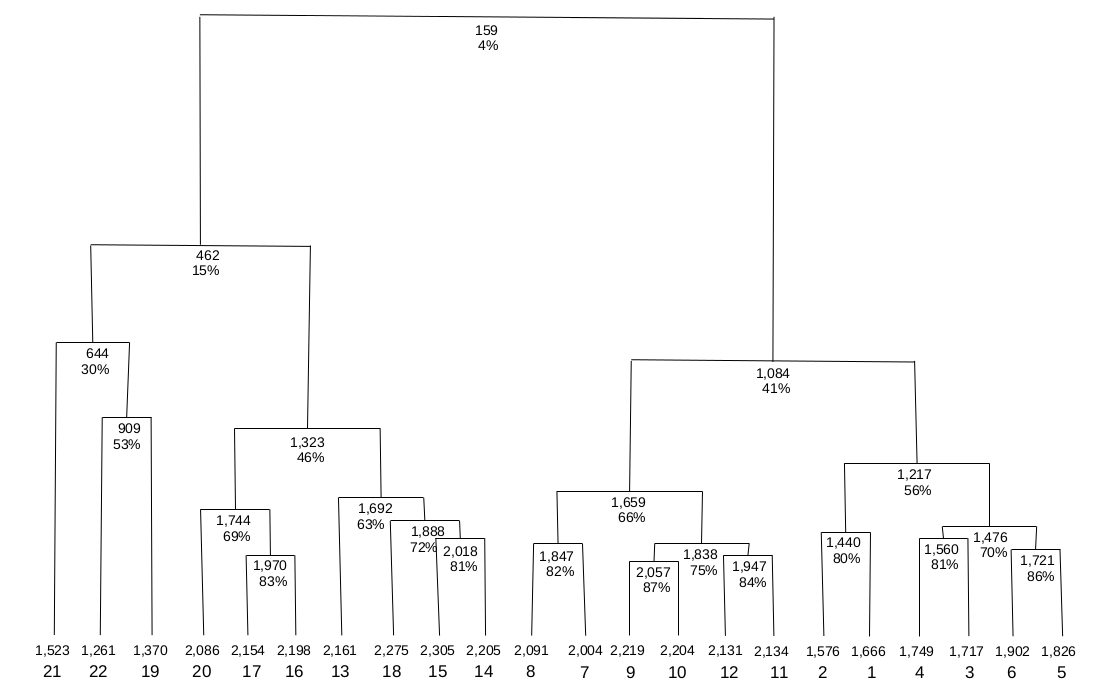}\hfill
\includegraphics[width=0.5\linewidth]{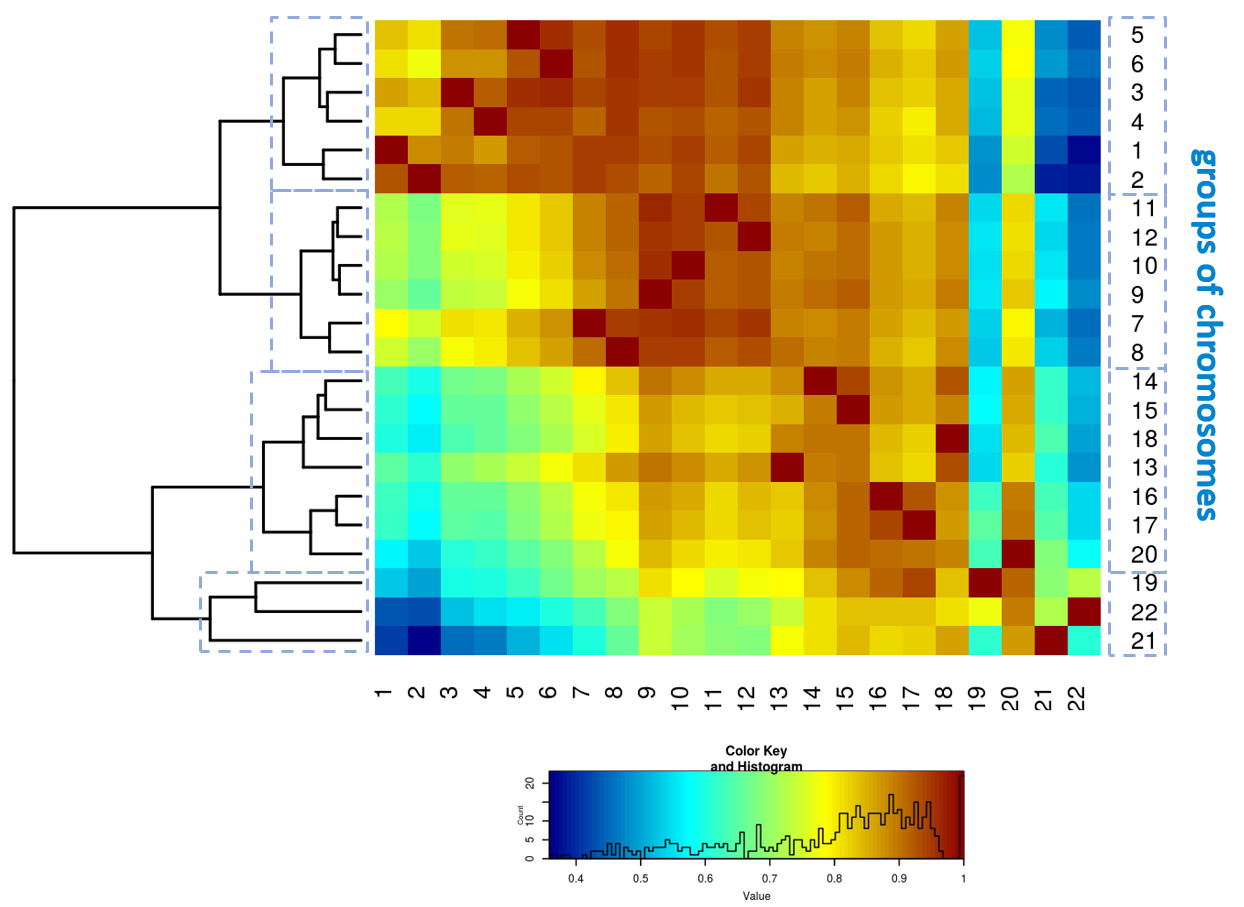}
\vspace{-0.4cm}\caption{\label{6mer} {\bf Human chromosome clusters.} Starting from the extracted seed5 non-reduced dictionary, the number (and percentage) of conserved examers in the pairwise intersection of human chromosomes is reported as dendrogram in top figure. The known similarity among chromosomes was found by our clustering, reported more in detail as heatmap in bottom figure. Four groups were identified, where dictionary intersection has a size comparable with that of single chromosomes. These (absolute and percentage) numbers are visible at the second level of the binary tree.}  \vspace{-0.3cm}
\end{figure}

We extracted dictionaries of examers on each single human chromosome, and from their pairwise intersections,rs in absolute and relative terms, we found interesting results, reported in Figure~\ref{6mer}, where four groups of chromosomes may be identified at the second level of the dendrogram, having cardinalities of dictionary intersection of the same order of that of the extracted dictionary from each single chromosomes (see leaves of the dendogram). Our dictionary based method was then capable to discriminate by structure similarity the following clusters of human chromosomes: $\{5,6,3,4,1,2\}$, $\{11,12,10,9,7,8\}$,$\{14,15,18,13,16,17,20\}$,$\{19,22,21\}$.

If we look back to Figure~\ref{elong}, we understand that prefixes of other words have to be kept in our dictionary, as {\it roots and morphemes of the genomic language}. As discussed in previous section, indeed, when starting by seeds longer than the first local maximum (of the word random deviation), we miss all the corresponding (shorter) significant words, obtained by starting with shorter seeds. On the other hand, here we point out that the word extraction to form the language must not depend on the single point in the genome where the seed is located, then we need to discard all suffixes from a dictionary (even in natural languages, suffixes are not proper words), as comprehensible by looking at Figure~\ref{elong} (b) and (c). Same argumentation holds for proper substrings, which may be generated by advancing the initial position of a seed, thus suffixes and substrings have been considered a computation bias and were both discarded. This is the motivation for a final filtering work on our dictionaries, to obtain the data reported in Table~\ref{22}, where suffixes have been isolated in order to form suffix-free dictionaries. 
\begin{table}[htbp]  \vspace{-0.4cm} \begin{center} 
\scalebox{0.7}{\begin{tabular}{r|r|r|r|r|r|r|r} \scriptsize
&  \multicolumn{3}{|c|}{suffixes} & & \multicolumn{3}{|c}{non-suffixes}\\
\hline
$k$	&	$|W_k|$ & $cov(G,W_k)$ & $avg(covp(G,W_k))$ & & $|W_k|$ & $cov(G,W_k)$ & $avg(covp(G,W_k))$ \\
\hline
6	&	1,121	&	0.749	&	2.114	&		&	545	&	0.492	&	1.528	\\
7	&	486	&	0.325	&	1.307	&		&	{\bf 1,824}	&	0.706	&	{\bf 2.218}	\\
8	&	47	&	0.017	&	1.084	&		&	546	&	0.129	&	1.482	\\
9	&	18	&	0.004	&	1.134	&		&	793	&	0.060	&	1.829	\\
10	&	10	&	0.001	&	1.015	&		&	2,130	&	0.020	&	1.180	\\
11	&	1	&	0.000	&	1.000	&		&	{\bf 2,114}	&	0.013	&	{\bf 1.783}	\\
12	&		&		&		&		&	1,363	&	0.009	&	1.181	\\
13	&		&		&		&		&	579	&	0.007	&	{\bf 1.377}	\\
14	&		&		&		&		&	145	&	0.006	&	1.224	\\
15	&		&		&		&		&	33	&	0.005	&	1.174	\\
16	&		&		&		&		&	17	&	0.002	&	1.254	\\
17	&		&		&		&		&	6	&	0.001	&	{\bf 1.396}	\\
18	&		&		&		&		&	1	&	0.001	&	1.000	\\
\hline
All	&	1,683	&	0.867	&	2.344	&		&	10,096	&	0.923	&	2.928	\\
\hline
\end{tabular}} \end{center} \vspace{-0.2cm}
\caption{\label{22} Dictionaries extracted by RDD-algorithm, from human chromosome 1, by starting from seeds 5 long, thus all seeds from $D_5(G)$. Dictionary sizes, sequence and (average) positional coverage values are reported, grouped for word lengths starting from 6. First three columns count the suffix words, while last three columns count non-suffix words (bold numbers denote the local maxima in the column).}\vspace{-0.9cm}
\end{table}

%\begin{table}[h] \begin{center}
%\begin{tabular}{r|r|r|r|r|r|r|r}
%&  \multicolumn{3}{|c|}{sub included} & & %\multicolumn{3}{|c}{non-subincluded}\\
%\hline
%$k$	&	$|W_k|$ & $cov(G,W_k)$ & $avg(covp(G,W_k))$ & & $|W_k|$ & %$cov(G,W_k)$ & $avg(covp(G,W_k))$ \\
%\hline
%6	&	1,632	&	0.841	&	2.728	&		&	34	&	0.039	&	%1.057	\\
%7	&	1,446	&	0.646	&	1.922	&		&	864	&	{\bf 0.464}	&	%{\bf 0.615}	\\
%8	&	186	&	0.074	&	1.297	&		&	407	&	0.096	&	1.327	%\\
%9	&	71	&	0.016	&	1.288	&		&	740	&	0.054	&	{\bf %1.740}	\\
%10	&	63	&	0.010	&	1.138	&		&	2,077	&	0.012	&	%1.073	\\
%11	&	22	&	0.007	&	{\bf 1.320}	&		&	{\bf 2,093}	&	0.006	%&	{\bf 2.269}	\\
%12	&	13	&	0.006	&	1.228	&		&	1,350	&	0.003	&	%1.093	\\
%13	&	11	&	0.005	&	{\bf 1.459}	&		&	568	&	0.001	&	%1.012	\\
%14	&	8	&	0.006	&	1.253	&		&	137	&	0.001	&	1.055	%\\
%15	&	3	&	0.003	&	1.134	&		&	30	&	{\bf 0.003}	&	%1.000	\\
%16	&	2	&	0.001	&	1.502	&		&	15	&	0.001	&	1.000	%\\
%17	&	2	&	0.001	&	1.768	&		&	4	&	0.001	&	1.000	%\\
%18	&		&		&		&		&	1	&	0.001	&	1.000	\\
%\hline
%All	&	3,459	&	0.979	&	3.778	&		&	8,320	&	0.587	%&	1.767	\\
%\hline
%\end{tabular} \end{center}
%\caption{\label{23} Dictionaries extracted by RDD-algorithm, from human %chromosome 1, by starting from $D_5(G)$ as seed set. Dictionary %cardinalities, sequence and (average) positional coverage values are %reported, grouped for word lengths starting from 6. First three columns %give the amount of infixes, while last three columns give the cardinality %of corresponding infix-free dictionaries (bold numbers denote the local %maxima in the column).}
%\end{table} 

A refined result from the application of RDD-based extraction algorithm on human chromosomes is a {\it reduced dictionary} generated by the suffix- and infix-free examers starting from seed length 5. This final selection of words turned out to include several well known biological sequences. We developed an empirical validation protocol for dictionaries in order to evidence its inclusion of biologically annotated regions, such as transcripts, lincRNA (long intergenic non coding RNA), CpG islands (often occurring close to the TSS, then overlapping some trascripts), sno miRNA (small nucleolar microRNA), TFBS (Transcription Factor Binding Sites), enhancers (of lengths from 200 to 2000) and other regulatory elements. For space limitation, we omit here to report these results.

The {\it non-reduced dictionary} instead is the dictionary extracted by the RDD algorithm without any suffix or infix filtering. In particular, the non-reduced dictionary of examers obtained by the algorithm from seeds long five was here employed to cluster all human chromosomes, as in Figure~\ref{6mer}, where known similarities between human chromosomes were confirmed by reverse engineering of our method. However, surprisingly, all chromosomes share very few examers (159 are common to all, over the 1,666 extracted words) which we exhibit as informative conserved sequences, a sort of product by evolution selection, to be further analyzed for their biological characterization. 

\section{Discussion}\label{final}

\begin{figure}[h]
\vspace{-0.2cm}\begin{center}
\includegraphics[width=0.8\linewidth]{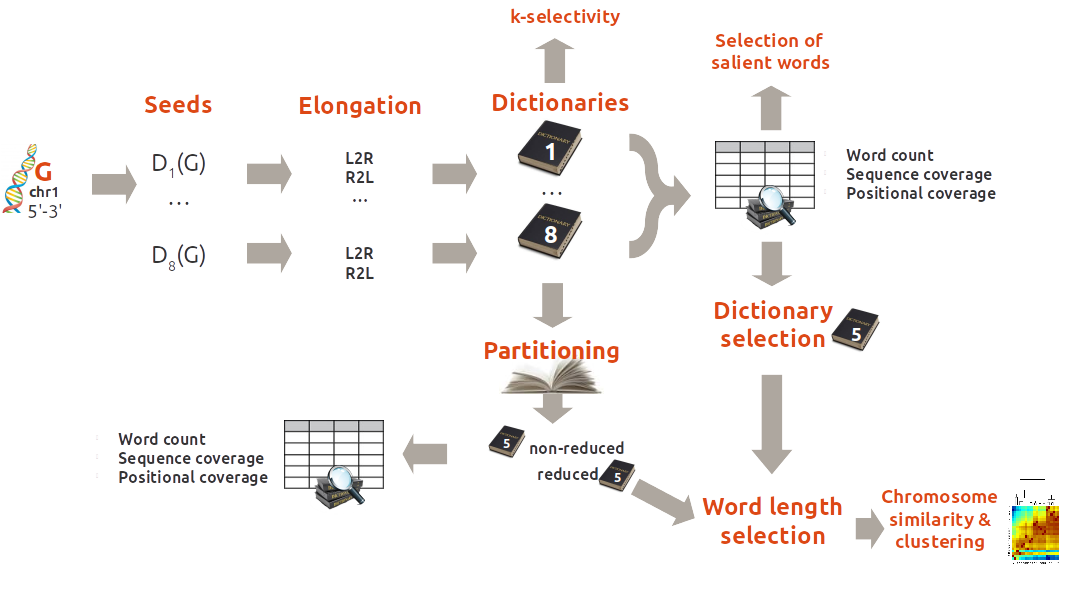}
\end{center}
\vspace{-0.8cm}
\caption{\label{meth} RDD based algorithm extracts a certain number of dictionaries, by starting from a given genomic sequence and by applying a word two-direction elongation to all seeds of initial length $k_0$ (that is, to all $k_0$-mers). The union of such dictionaries may be analyzed, filtered, partitioned, according to different strategies (Word Length Distributions, coverage properties, etc). Reduced dictionaries are suffix-free and infix-free, and provide possible genomic languages. Non-reduced dictionaries of examers have been employed to identify chromosomes similarity and clustering.} \vspace{-0.3cm}
\end{figure}

Given a genome we extract a specific set of its factors which represent the building blocks, or semantic units, of a dictionary significant for the genome language. In this work we have described an information theoretical methodology to extract relatively small genomic dictionaries, which have good properties in terms of genome coverage, illustrated in Figure~\ref{meth}.

Namely, three methods were presented. One from the literature, introduced in~\cite{carpena2009level}, which was our starting point in terms of basic ideas, the second method is a variant of this, called V-algorithm, more efficient and appropriate to extract genomic dictionaries, and finally, our RDD based W-algorithm, which originally combines a criterion of anti-randomness with a criterion of elongation of seeds to select variable length factors.

More precisely, both the V- and W-algorithms are based on concepts of word elongation and anti-randomness, though in a complementary way. Both consider word elongations significant if the distribution is far from random according to a given parameter, which is C for the V-algorithm and the value of r in equation (1) for the W-algorithm. Parameter C measures the clustering of word occurrences, then its difference from random increases for words densely distributed (in average), while r measures the difference between the average distance of word recurrence (i.e., the distance between two occurrences) and that expected in a Poisson process. This is the suitable distribution of rare events, such as word occurrences in (long) genomes~\cite{castellini}. We may roughly say that the V-algorithm is based on the high word density and the W-algorithm is based on the non-random word recurrence.

In section~\ref{Valg} performance results are shown in a comparison between the V-algorithm and the original one introduced in~\cite{carpena2009level}.
Here we may add that both applied methodologies show a weakness in elongating words: often they are not able to extend seed words. On the other hand, when they succeed in word elongation, they extract very long words (up to $1,941$ and $1,388$, respectively, the original and the variant algorithm) which look away from the concept of language unit. However, a difference between the two versions is that, the V-algorithm selects shorter words (lengths from $6$ to $8$-$10$) and discard them by removing redundant prefixing words, while the same does not happen for the original version. This confirms that the variant is more stable than the original one, indeed, by varying the seed length $k_0$, the same semantic prefixes are kept.

The point of our approach is to find best seed lengths to optimize the performance of the extraction algorithm, and eventually appropriate word lengths as good dictionaries to investigate. Best performance is obtained by producing relatively small dictionaries with both sequence and average positional coverage as close as possible to one. In section~\ref{Walg} we have exhibited dictionaries extracted by the W-algorithm, that own an internal bimodal trend and the property such that dictionaries starting from shorter seeds are contained (relatively to words of the same length) by those obtained starting by longer ones (see Tables~\ref{1WLD},~\ref{2WLD}). Preferred seed lengths emerge, from an observation of sequence and positional genome coverage (see Tables~\ref{Seq},~\ref{Pos}), that provide a better coverage. Dictionaries of examers were identified to reveal a clear similarity pattern for human chromosomes. Then genomic dictionaries were extracted and filtered (by elimination of suffixes and substrings), which resulted in languages composed by roots and morphemes, including known biological sequences.

Other numerical analysis were performed on single chromosomes. As an example, we considered comparisons with examers extracted by other seeds length, or with examers composing extracted longer words. Namely, for human chromosome 22 we analyzed all the examers contained in the 40 words of length greater than 17, extracted by the algorithm starting from seed 8, to check if they have intersection with the 1,261 examers we see in Figure~\ref{6mer}. The surprising result is that 139 over the 256 (i.e., a majority of) examers extracted from the 40 words do not belong to the dictionary of 1,261 examers (extracted by seed 5). Longer words obtained with longer seeds are composed by generally different examers than those in the dictionary with shorter seeds, and this is promising for the validity of our method, producing almost infix-free dictionaries.

For future developments of our methodology, we will consider further concepts useful to evaluate dictionaries, and extend our extraction methodology to patterns of words. For example, it is reasonable to assume that selected words have to present a low percentage of possible anagrams that effectively appear within the genomic sequence.

\subsection{Biological validation}

As a further validation of our method, it turned out that extracted informational dictionaries include several well known biological sequences. We developed an empirical validation protocol for dictionaries that helped us to find evidence of several notices, which we report here for human chromosome 22, especially for dictionaries D of examers obtained by seeds with a length in the range 1-5.

To collect some of the main biologically annotated regions, into a dictionary we called A, we used UCSC Genome Browser~[\cite{browser}]. Our dictionary A includes: transcripts (from RefSeq~[\cite{refseq}]), lincRNA (long intergenic non coding RNA), CpG islands (often occurring close to the TSS, then overlapping some trascripts), sno miRNA (small nucleolar microRNA), TFBS (Transcription Factor Binding Sites), enhancers (of lengths from 200 to 2000) and regulatory elements (from ORegAnno database: Open Regulatory Annotation). Class of lincRNA is included in that of lncRNAs (long non-coding RNA), having several biological functions, together with ncRNA (non coding RNAs shorter than 200bp), such as transcription, splicing, traslation, cellular cycle, apoptosis. Elements lincRNA show a tissue specific expression~[\cite{linc}] and have an average length of 1000 bases, then shorter than that of protein coding transcripts (in average 2900 long).

In chromosome 22 (130,481,394 bp long) the above annotated regions count 23,209,047 bases. However, since they share overlapping words of the chromosome (specifically, 7.1\% bases of the annotated regions), the dictionary A involves 21,567,860 of the whole chomosome22. Therefore, A has chromosome coverage equal to 61.8\% while 38.2\% remains the non-annotated region. By extraction of dictionary D, there are 14 words which occur only in the non-annotated region, and more in general, there are 114 words having more than 80\% of their occurrences in such a non-annotated region. This is notable, if we consider that this threshold is double percentage than expected, and we presume an informative value for these specific words.

In Table~\ref{ann} some annotated words of A are reported (in the first column) with their related coverage of the dictionary of examers extracted from seed1. We may notice that the words of the dictionary cover 25.80\% of the transcripts, 24.9\% of lincRNAs, 25.77\% of the whole chromosome, and 28.09\% of the exons. Indeed, if we separate data on introns and exons (as in Table~\ref{in-ex}), protein coding (25.8\%) and non-coding transcripts (25.7\%) ratios do not change. From data in Table~\ref{ann} we conclude that the dictionary extracted by seed1 is present as expected in all the annotated regions we considered, with the exeption of CpG islands, where it is 77\% more present than expected, however this is due to the high content of GC in short words (which indeed exhibit higher coverage parameters).

The CpG islands are the most covered region by the dictionary (0.491), however, this does not mean they constitute the most informative region. Indeed, words from the extracted dictionary, of length from 4 to 6, have a high number of couples G+C (between 0.64 and 0.81), even if they were mostly extracted outside of the CpG islands (which cover only the 2\% of the chromosome). Words of length 5 and 6, in particular, exhibit a large chromosome coverage, and mainly contribute to compute the corresponding value in Table~\ref{ann}.
\begin{table} \begin{center}
\begin{tabular}{r|r|}
Annotated words & $cov(A,D)$ \\
\hline
trascripts & 0.258 \\
lincRNA & 0.249 \\
CpG islands & 0.491 \\
sno miRNA & 0.248 \\
TFBS & 0.287 \\
enhancers & 0.309 \\
regulatory elements & 0.264 
\end{tabular}
\end{center}
\caption{\label{ann} Annotated words coverage by the extracted dictionary from chromosome 22, starting by seed1.} 
\end{table}

Chromosome 22 reaches a maximum positional coverage (equal to 0.52) with examers obtained by elongation of seeds long from 1 to 5 (of course, no other length of seed would produce examers, then we may say they are examers produced by RDD-algorithm). Coverage of such a dictionary over exons and introns of chromosome 22 may be seen in Table~\ref{in-ex}.
\begin{table} \begin{center}
\begin{tabular}{r|r|r|r|}
Seed (amount of examers) & $cov(exons,D)$ & $cov(introns,D)$ & $\frac{cov(exons;D)} 
{cov(introns;D)}$\\
\hline
Seed1 (63) & 0.1503 & 0.1370 & 1.10 \\
Seed2 (349-63=286) & 0.5246 & 0.4930 & 1.06 \\
Seed3 (517-349=168) & 0.3444 & 0.3205 & 1.07 \\
Seed4 (995-517=478) & 0.6434 & 0.6275 & 1.03 \\
Seed5 (1,261-995=266) & 0.4169 & 0.4010 & 1.04 
\end{tabular}
\end{center} \caption{\label{in-ex} New examers are added with the row index (and with the seed length) and their positional coverage is measured over exons and introns. The values represent the percentage of exons and introns (in the transcripts) which is covered by the dictionary of examers extracted by seeds with length in the range~1-5.} 
\end{table}

Finally, we verified that words (of the examers obtained by seed1) which occur (about 200 times) across transcripts and exons coincide with {\it known splicing recognition sites}: they are examers (G{\bf CAG$\mid$G}C, {\bf CAG$\mid$G}GA e {\bf CAG$\mid$G}GC, consensus sequence is denoted in bold). Other examers of the dictionary occur at most 121 times across exons. Transcripts coverage and chromosome coverage turn out to coincide for fixed length extracted dictionaries (subdictionaries partitioned by word length, as in previous WLD tables), the same holds for lincRNA. If we focus on words long 4, we found a couple of them (CG$\mid$GT, G$\mid$GAG) occurring 3 times the others (600 versus a maximum of 200) which coincide with known sequences across exons and introns.

%The evaluation of extracted dictionaries is based on the degree of the whole sequence coverage, the degree of single genomic position coverage, and on the number and types of known biologically significant words they contain.

%Computational genome analyses where specific informational concepts are massively investigated can unravel the internal logic of genome organization, where rigorous mechanisms and chance are mixed together to achieve the main features that are proper of living organisms~\cite{booklet}. The importance of ``good'' dictionaries has a number of implications for future works on computational genome analysis. %In fact, distributions are crucial in genome analysis, and many important genomic distributions are based on dictionaries.

%We intend to continue this investigation on multi-thread analysis, by comparing (and finding similarities) among different strains and organisms.

\section{Future work}

More in general, one of the future possible developments of the present approach is the definition of genomic entropic divergences, as a way for comparing genomes and for discovering genomic differences and similarities at a global structural level. Indeed, when two dictionary $D_1$ and $D_2$ are extracted from two genomes $G_1$ and $G_2$, it is interesting to associate two probability distributions related to these dictionaries. An idea (see~\cite{prof3}) is that of defining a common dictionary $D$ formed by the longest common prefixes of $D_1$ and $D_2$. Their  intersection $Pref(D_1) \cap Pref(D_2)$, after the elimination of words which are prefixes of others, is a  good candidate for such a dictionary $D$, because two distributions may be defined for the two genomes, by $p_1(\alpha) = \sum_{\alpha\beta \in D_1} p(\alpha\beta)$ and $p_2(\alpha) = \sum_{\alpha\beta \in D_2} p(\alpha\beta)$, respectively, where in both cases, $p(\alpha\beta)$ is the frequency of the string $\alpha\beta$ in $G_1$ or in $G_2$, respectively. The entropic divergence between $p_1$ and $p_2$ is a measure of their whole difference, and the more the two dictionaries are expressive for the two genomes, the more this measure is accurate. 

%\bibliographystyle{unsrt}
%\bibliography{biblio}

\end{document}